\documentstyle[natbib,emulateapj]{article}
        
\def\etal{{\it et~al. }}

\begin{document}

\title{The Measurement of Disk Ellipticity in Nearby Spiral Galaxies}

\author{David R. Andersen}
\affil{Department of Astronomy and Astrophysics,
Pennsylvania State University, 525 Davey Lab, University Park, PA
16802; andersen@astro.psu.edu.}

\author{Matthew A. Bershady, Linda S. Sparke, John S. Gallagher, III,
Eric M. Wilcots} \affil{Department of Astronomy, University of
Wisconsin, 475 N Charter Street, Madison, WI 53706;
mab@astro.wisc.edu, sparke@astro.wisc.edu, jsg@astro.wisc.edu,
ewilcots@astro.wisc.edu.}

\begin{abstract}

We have measured the intrinsic disk ellipticity for 7 nearby, nearly
face--on spiral galaxies by combining Densepak integral-field
spectroscopy with $I$-band imaging from the WIYN telescope.
Initially assuming an axisymmetric model, we determine kinematic
inclinations and position angles from H$\alpha$ velocity fields, and
photometric axis ratios and position angles from imaging data. 
We interpret the observed
disparities between kinematic and photometric disk parameters in terms
of an intrinsic non-zero ellipticity, $\epsilon$.  The
mean ellipticity of our sample is 0.05.  If the majority of disk
galaxies have such intrinsic axis ratios, this would account for
roughly 50\% of the scatter in the Tully--Fisher relation. This
result, in turn, places tighter constraints on other sources of
scatter in this relation, the most astrophysically compelling of which
is galaxy mass-to-light ratios.

\end{abstract}

\keywords{galaxies: structure -- galaxies: kinematics}
              
\section{Introduction}

A large fraction of galaxy disks appears to be non-axisymmetric.
Photometric studies of 
face--on galaxies find the light distribution in 30--50\% of
spirals is lopsided \citep[$m=1$ distortions;][]{zaritsky97,
rudnick98} or
has higher order distortions
\citep[$m\geq2$;][]{rix95,kornreich98,conselice00}. A similar fraction
of galaxies has lopsided HI distributions or kinematic asymmetries
\citep{baldwin80,bosma81b,richter94,haynes98,swaters99}.

Disk ellipticity
($m=2$) may also be common.  Few disk galaxies appear 
round on the sky; the distribution of apparent axis ratios is best
modeled by randomly oriented disks with intrinsic ellipticities
$\epsilon\sim0.1$
\citep{binney81,grosbol85,huizinga92,lambas92}. These
statistical studies constrain mean ellipticity, but
do not probe the intrinsic ellipticity of individual galaxies.

If galaxy disks are elongated, errors in photometric
estimates of inclinations and position angles (PAs) will be
introduced, thereby contributing to the scatter in the relationship
between galaxy luminosity and rotation speed \citep{tully77}.
\citet{franx92} used a specific, non-rotating velocity field model to
show that if $\epsilon=0.1$ on average, systematic errors could
account for all the observed scatter in the Tully--Fisher (hereafter,
TF) relation.  However, the true contribution to this scatter depends
on the distribution of intrinsic axis ratios. Identifying the 
individual ellipticities of galaxies would
limit other sources of
astrophysical scatter in the TF relation and reduce scatter.

Kinematic maps can be used to estimate ellipticity for individual galaxy
disks. One
approach involves fitting such maps with elliptic orbits, from which
a quantity related to the ellipticity of the potential can be estimated:
$\epsilon_{\rm pot} \sin 2\phi$. This quantity
varies from 0.001 -- 0.07 for a sample of 9 galaxies
\citep{schoenmakers97,schoenmakers99}. Unfortunately, the 
phase angle $\phi$ is not directly measurable. 
Such studies have been rare because they require
observationally expensive, high resolution, high signal-to-noise
velocity field maps \citep[e.g.][]{teuben86,kornreich00}.

Here we develop a new, efficient method of estimating intrinsic galaxy
disk ellipticities using optical kinematic maps and photometric
indices. Determining unique solutions for $\epsilon$ and $\phi$
requires measurements of both kinematic and photometric
inclination and PAs for nearly face--on
spiral disks. When the disk major axis is rotated by a phase angle
$\phi$ away from the line of nodes, the kinematic and photometric axes
will appear misaligned only if $\cos i < 1-\epsilon$. If
$\epsilon\approx 0.1$, as suggested by earlier studies, substantial
misalignment of kinematic and photometric PAs ($>30^\circ$) will
occur only for galaxies with $i < 30^\circ$.
To demonstrate our method we present the
detailed analysis of WIYN integral field H$\alpha$ velocity fields and $I$
band images for 
UGC 4380, and measurements of $\epsilon$ for an
additional 6, nearly face--on galaxies based on the same method.

\section{Sample and Observations}

Our sample of seven galaxies is a subset of 69 selected from
the {\sl Principal Galaxy Catalog} \citep[PGC;][]{paturel97}
in regions of low Galactic extinction with normal surface-brightnesses,
axis ratios corresponding to $<30^\circ$ inclination
for a circular disk, diameters D$_{25}$ between 0.75 and 1.5
arcminutes (to facilitate spectroscopic follow-up), and Hubble types
between Sb and Sc \citep[see][]{andersen01}. We excluded 
galaxies with obvious bars, 
clearly visible non-axisymmetric structure, and 
foreground stars within 2-4 disk scale lengths using
the Second Palomar Observatory Sky
Survey.

We used the WIYN S2KB imager (a 2048$^2$ CCD with 0.195 arcsec/pixel)
to obtain deep $I$-band images of our 7 targets between May 10-14,
1999 with seeing 
of $\sim$1.1 arcseconds FWHM, or 10\% of 1 scale length.  We reduced
these images using standard procedures, yielding sufficient
signal-to-noise to establish galaxy isophotes to $\sim4.5$ scale
lengths.  We used flux-calibrated, $R$-band magnitudes obtained with
the IGI/TK4 imager on the McDonald Observatory 107'' telescope to
check that these galaxies had close to ``Freeman'' disks ($B$-band
central disk surface brightness of 21.7 mag arcsec$^{-2}$).  For
example, assuming $B-R=1.15$ for the Sc-type galaxy UGC 4380, we find
$\mu_0(B)=21.5$ mag arcsec$^{-2}$.

Spectral images of our 7 targets were obtained using Densepak on the
WIYN telescope from portions of runs on May 20-21, 1998, January
20-22, 1999, and March 27-28, 1999. Densepak is a fiber-optic array
used for integral field spectroscopy containing 86 active,
3-arcsecond fibers (separated by 4 arcseconds)
arranged in a $7\times 13$ staggered grid subtending an area of $30
\times 45$ arcseconds \citep{barden98}. Four additional ``sky'' fibers
are spaced around this grid roughly one arcminute
from the grid center. Densepak feeds the WIYN Bench Spectrograph,
which we used with an echelle grating to cover $6600{\rm \AA} <
\lambda\lambda < 7000$\AA~with a dispersion of 0.19~\AA~per pixel and
a FWHM spectral resolution of 0.51~\AA.  Basic spectral reductions
were done using the NOAO {\sl IRAF} package {\it
dohydra}. Thorium-Argon lamps were used for wavelength calibration.

Since galaxy rotation curves peak at roughly two photometric scale
lengths \citep{courteau99,willick99}, we used Densepak to map out to
$\sim 3$ scale lengths per galaxy (typically
$\sim 30$ arcsec). For UGC 4380, which has a $I$-band
scale length of 9.7 arcseconds, two Densepak pointings were
required to cover a square of roughly 45$\times45$ square arcseconds. 
Two 30-minute exposures were made at each Densepak position
to facilitate cosmic ray removal. The H$\alpha$ line flux was
typically measured at a signal to noise of $\sim10$ at the edge of our
observed field. 

\section{Measurements}

To measure photometric axis ratios and PAs we use the STSDAS ISOPHOTE
package {\it ellipse} on images with stars masked. Figure 1 shows the
radial dependence of PA, apparent ellipticity (1-$b/a$), and an
azimuthal variance statistic for UGC 4380. Inside 29 arcseconds,
the variance statistic is greater than unity showing that
spiral arm structure affects the ellipse fits.
Hence we measure the photometric axis ratio
and PA at radii greater than 29 arcseconds, where the axis ratio and
PA are constant, consistent with our estimate that spiral structure
has a minimal impact.

To determine kinematic inclination and PAs, we first produce H$\alpha$
emission line 
maps from the 86 field-flattened and sky-subtracted spectra 
from each pointing. We measure velocity centroids by
fitting a Gaussian line profile to each H$\alpha$ line, and 
assign a spatial position based on the fiber geometry of Densepak and
the offsets used to make multiple pointings. Figure 2 shows a
polynomial surface fit to these discrete velocity measurements for UGC
4380. UGC 4380 clearly is an inclined galaxy with a
center coincident with the photometric center. The
formally well-determined PA and peak rotation velocity (Table 1) agree
with visual inspection of the data (Figure 2). Single dish HI
observations of UGC 4380 list a $W_{50}$ velocity widths of 118 
km/s \citep{haynes88}.  The $W_{50}$ width measured
from our Densepak spectra is also 118 km/s, implying we
observed the peak of the rotation curve.

To extract the PA, peak velocity and inclination,
we fit the velocities expected at each fiber position
using a simple model consisting of concentric and coplanar
circular orbits with $V(r) = V_c \tanh (r/h)$, where the terminal
circular velocity $V_c$ and $h$ are free parameters.  The other free
variables are inclination, PA, center, and central velocity.  
Our velocity fields exhibit only small deviations from this simple 
model, as illustrated for UGC 4380 in the middle panel of Figure
2. The best fitting model was determined from $\chi^2$ minimization
(downhill simplex method) based on comparing the measured velocity
centroids, fiber by fiber with the smooth model velocity field sampled
with the Densepak footprint. The standard deviation in the fit
residuals is 4.3 km/s.  More elaborate radial velocity functions do
not provide smaller residuals yet have more independent variables.
Formal confidence limits (CL) were placed on these quantities by
determining surfaces of constant $\chi^2$.  Table 1 contains our
measurements of axis ratio, photometric PA and kinematic inclination
and PA.  In UGC 4380, the photometric and kinematic PA differ by
$9.5^\circ\pm3.9^\circ$ (68\% CL).  This disk is unlikely to be
intrinsically circular.

\section{Analysis and Results}

We can estimate the ellipticity required to produce the above level of
discrepancy by modeling the observed galaxy as a disk with intrinsic,
photometric ellipticity $\epsilon$ (right panel, Figure 2) but
circular orbits (justified below). The
vector describing the ellipse in the galaxy plane
is $\vec r~^\prime=
[-(1-\epsilon)\sin\theta ,\cos\theta ]$, where $\theta=0$ is the major
axis.
To project the ellipse onto the (observed) sky plane, it is inclined by
an angle $i$ about the line of nodes at an angle $\phi$ from
the intrinsic major axis.
In the plane of the sky, the ellipse describing the galaxy isophotes
is given by a transformation matrix involving $\theta$, $i$,
$\epsilon$, and $\phi$.

\citet{franx92} found that for orbits in a flat but elliptic disk, 
fitting the
velocity field with tilted circular rings yields the correct disk
orientation to first order in ellipticity.  Accordingly, we take the
kinematic inclination and PAs to represent the true inclination and PA
of the disk; $\epsilon$ and $\phi$ can then be determined given
measurements of the apparent photometric axis ratio (${b/a}$), the
kinematic inclination ($i$), and the difference between
photometric and kinematic PAs ($\psi$).

When the true inclination $i$ is closer to face--on than
the photometric inclination, as is the case for UGC 4380, the galaxy
must be at least as flattened as $\epsilon \geq 1-{b\over a}\cos
i^{-1}$. The intrinsic flattening must be even larger if the true
major axis does not lie in the plane of the sky.  This lower limit for
UGC 4380, $\epsilon > 0.06\pm 0.03$, is inconsistent with a purely
circular disk.  By utilizing the $9.5^\circ\pm3.9^\circ$ misalignment
of the photometric and kinematic PAs, a better estimate of the
ellipticity can be obtained.

In general, three equations relate the three observables ${b/a}$, $i$,
and $\psi$ to the three unknowns $\epsilon$, $\phi$, and the angle
$\theta$ at the apparent major axis.  Using the Newton--Raphson method
for nonlinear equations, we find $\epsilon = 0.07^{+0.08}_{-0.06}$
(99\% CL) for UGC 4380, a solution {\it in}consistent with a circular
disk. Derived values for the seven galaxies, plotted in Figure 4 and
listed in Table 1, range from $\epsilon = 0.02$ to 0.20.  Three
galaxies (UGC 4380, NGC 2794, UGC 5274) are inconsistent
with having circular disks at the 99\% CL; two galaxies (NGC 3890, NGC 5123)
have ellipticities inconsistent with circular disks at the 95\% CL.  Only two
galaxies (UGC 7208 and UGC 10436) are consistent with having circular
disks within their 68\% CL. The galaxies with the highest derived
ellipticity $\epsilon$ have $\phi$ near 90$^\circ$; the line of nodes
is almost perpendicular to the true major axis. This is consistent with
our selection of round, apparently face-on systems.
The two galaxies with $\epsilon>0.1$, NGC 2794
and UGC 5274, both show evidence for faint, interacting companions; NGC
2794 has an AGN.

Finally, we have checked that potential sources of systematic errors
-- lopsidedness, spiral structure, and warps -- do {\it not} affect
our ellipticity measurements appreciably. Table 1
contains the $\langle A_1\rangle$ amplitude of the $m=1$ component of the 
Fourier expansion
of the light profile as defined by \citep{zaritsky97}.
With the exception of NGC~5123
these galaxies are not significantly lopsided ($\langle A_1\rangle <0.2$),
nor does $\langle A_1\rangle$ correlate with our derived values
for $\psi$ or $\epsilon$.
None of these galaxies show any sign of kinematic
asymmetry. A 180$^\circ$-rotational asymmetry measure, defined as $
A_{180}=|\sum (V_{obs} + V_{180})|/ |2 \sum V_{obs}|$, akin to the
photometric asymmetry parameter of \citet{conselice00}, yields close
to null values for all. 

Strong spiral structure drives photometric PAs
to change with radius $R$ at a rate of $\partial {\rm PA}/\partial\log R =
\cot\theta_p$, where $\theta_p$ is the pitch angle of the arms.
A warp in the disk also would be
manifest as a twisting PA with radius. As
discussed in \S3, we make our photometric measurements 
between 3--4 scale-lengths where
we find the photometric PA and axis ratio are constant, and our azimuthal
variance statistic (Figure 1) corroborates that spiral structure is no
longer a dominant photometric effect. The velocity fields exhibit
no residual structure correlating with radius or azimuth, despite
the fact that these measurements are {\it within} 3 scale-lengths (in
contrast to our photometric measurements). 
We find no evidence for twisting PAs in the velocity
fields which is not surprising since even galaxies with strong outer
warps usually have planar
HI distributions within 3 scale lengths \citep{briggs90}.

\section{Summary and Discussion}

We have demonstrated that high-quality H$\alpha$ velocity maps and $I$
band images of nearly face--on galaxies can be used to {\it exclude
the hypothesis that galaxy disks are intrinsically free of $m=2$
(elliptic) distortions.} Our method for estimating the deviation from
circularity suggests UGC 4380 has an intrinsic ellipticity of
$\sim$7\%. This is one of the first unambiguous detections of disk
ellipticity for an individual galaxy. If disks are intrinsically
elliptic, the photometric and kinematic axes will be 
misaligned in general, and the inclinations derived from 
isophote shapes will
differ from the kinematic inclinations.  Deviations are particularly
large for face--on galaxies.  WIYN/Densepak echelle spectroscopy and
optical imaging are efficient means to estimate disk ellipticity for
large samples of such systems.

The intrinsic ellipticity for our 7 galaxies is
$\epsilon=0.05\pm0.01$ (error weighted mean and uncertainty), inconsistent 
with purely circular disks. Most of our targets are normal,
intermediate-type spirals, typical of those selected for TF
surveys. According to the \citet{franx92},
$\epsilon=0.05$ should produce $\sim$50\% of the observed TF
scatter in red and near-infrared bands.  The cause of disk
ellipticity is currently unclear, e.g., halo triaxiality or
non-uniform matter accretion could be responsible. If larger samples 
show disk ellipticity
at these levels, this implies variations in either disk mass fractions or
mass-to-light ratio of today's spiral galaxies must contribute under
0.1-0.2 mag dispersion in the luminosity of galaxies at a given
rotation speed.  High resolution cosmological
simulations \citep[e.g.][]{steinmetz99,bosch00} indicate that variance
in disk mass fraction should only induce scatter {\it along} the
TF relation, because halo contraction is greater for larger disk
masses. Assuming the remaining scatter in the observed TF relation is
dominated by differences in spiral galaxy mass-to-light ratios, this
modest variance places strong constraints on the formation histories
of such systems.
 
\acknowledgements We thank S. Barden at NOAO and D. Sawyer at WIYN
for making Densepak
happen. This research was supported by the UW-Madison Graduate
School, and NSF grants AST-9618849 and AST-9970780 (MAB).

\bibliographystyle{apj}

{\scriptsize
\begin{deluxetable}{llrrrrllllll}
\tablenum{1}
\tablewidth{0pt}
\tablecaption{}
\tablehead{
\colhead{ID} & 
\colhead{$b/a$} &
\colhead{PA$_{\rm phot}$} & 
\colhead{$i_{\rm kin}$} &
\colhead{PA$_{\rm kin}$} &
\colhead{$\psi$} &
\colhead{$\phi$} &
\colhead{$\epsilon$} &
\multicolumn{3}{c}{$\epsilon$ confidence limits} & 
\colhead{$\langle A_1\rangle$} \\ \cline{9-11}
\colhead{} & 
\colhead{} &
\colhead{(deg)} & 
\colhead{(deg)} & 
\colhead{(deg)} & 
\colhead{(deg)} & 
\colhead{(deg)} & 
\colhead{} &
\colhead{68\%} &
\colhead{95\%} &
\colhead{99\%} &
\colhead{} 
}
\startdata
UGC 4380 & 0.90$\pm0.02$  &  43$\pm4$ & 16.1$\pm2.3$ & 33.5$\pm1.8$  & 10$\pm9$  & 15$\pm15$ & 0.07 & {\small $^{+0.04}_{-0.04}$} & {\small $^{+0.06}_{-0.06}$} & {\small $^{+0.08}_{-0.06}$} & 0.06$\pm0.02$ \nl
NGC 2794 & 0.85$\pm0.04$  & 408$\pm3$ & 20.5$\pm1.4$ & 327.9$\pm0.9$ & 80$\pm3$ & 83$\pm4$  & 0.20 & {\small $^{+0.07}_{-0.07}$} & {\small $^{+0.11}_{-0.11}$} & {\small $^{+0.13}_{-0.13}$} & 0.04$\pm0.02$ \nl
UGC 5274 & 0.94$\pm0.02$  &  78$\pm7$ & 32.2$\pm3.8$ & 9.5$\pm1.4$   & 69$\pm7$ & 84$\pm5$  & 0.20 & {\small $^{+0.07}_{-0.06}$} & {\small $^{+0.12}_{-0.10}$} & {\small $^{+0.14}_{-0.11}$} & 0.07$\pm0.02$ \nl
NGC 3890 & 0.96$\pm0.02$  & 264$\pm16$ & 23.5$\pm1.8$ & 225.6$\pm1.4$ & 38$\pm16$ & 73$\pm13$ & 0.08 & {\small $^{+0.05}_{-0.04}$} & {\small $^{+0.07}_{-0.07}$} & {\small $^{+0.08}_{-0.08}$} & 0.13$\pm0.06$ \nl
UGC 7208 & 0.94$\pm0.01$  & 325$\pm9$ & 23.8$\pm2.3$ & 330.4$\pm1.2$ & -5.4$\pm9$ & 101$\pm90$ & 0.03 & {\small $^{+0.04}_{-0.03}$} & {\small $^{+0.06}_{-0.03}$} & {\small $^{+0.08}_{-0.03}$} & 0.12$\pm0.05$ \nl
NGC 5123 & 0.92$\pm0.01$  & 163$\pm4$ & 22.2$\pm1.1$ & 153.0$\pm0.5$ & 10$\pm4$ & 44$\pm27$ & 0.03 & {\small $^{+0.02}_{-0.02}$} & {\small $^{+0.04}_{-0.03}$} & {\small $^{+0.04}_{-0.03}$} & 0.25$\pm0.10$ \nl
UGC 10436& 0.87$\pm0.03$  & 270$\pm3$ & 30.6$\pm1.4$ & 267.1$\pm0.7$ & 3$\pm3$  & 64$\pm90$ & 0.02 & {\small $^{+0.06}_{-0.02}$} & {\small $^{+0.09}_{-0.02}$} & {\small $^{+0.11}_{-0.02}$} & 0.09$\pm0.05$ \nl
\enddata
\end{deluxetable}
}

\clearpage

\figcaption{The radial dependencies of photometric position angle (PA;
top panel), ellipticity (1-$b/a$; middle panel) of UGC 4380's $I$-band
isophotes. The bottom panel shows the azimuthal surface-brightness
variance around each isophotal ellipse, normalized by the expected
shot-noise (source noise, $\sigma_s$, and sky plus read-noise,
$\sigma_b$). This normalized variance is large ($>1$) where spiral
structure contributes to the overall variance budget. We measure the
PA and $b/a$ between 29 and 39 arcseconds where both the PA and axis
ratio remain constant, the azimuthal surface-brightness variance is
consistent with shot noise, and the signal to noise per pixel is greater
than one.}

\figcaption{The Densepak H$\alpha$ velocity field of UGC 4380 (left
panel) and the residuals between the velocity field and a simple model
(middle panel; see text). Both are smoothed and interpolated using a
polynomial surface.  Solid, heavy, and dashed lines are positive,
zero, and negative velocities, respectively, relative to the model
systemic velocity (left panel) or model velocity field (middle panel).
The dash-dotted lines (left panel) represent the isophotal annulus
determined from Figure 1 from which the photometric ${b/a}$ and PA are
derived (solid lines in annulus indicate photometric major and minor
axes).  The right, summary panel shows a schematic diagram of an
isophote for UGC 4380 assuming its disk has an intrinsic ellipticity
of $\epsilon=0.07$. The $x$-$y$ plane has been rotated 33.5$^\circ$
from North to match the kinematic position angle of UGC 4380. The
solid ellipse represents the true shape of the disk seen face-on; the
major axis lies at an angle $\phi = 15^\circ$ from the line of nodes
(the kinematic major, or y axis). The dashed ellipse represents the
apparent shape after inclining the elliptic disk by $i=16.1^\circ$;
it has an apparent axis ratio of ${b/a}=0.90$. The
kinematic and photometric PAs differ by $\psi=9.5^\circ$.}

\figcaption{Solutions for ellipticity $\epsilon$ and angle $\phi$ for
seven sample galaxies. The 68\% confidence limits (CL), shown as
contours, are derived from the estimated measurement errors on $b/a$,
$\psi$ and $i$ listed in Table 1. UGC 7208 (dark-dash) and UGC 10436
(light dash) are consistent with a circular disk within the
68\% CL.}

\clearpage

\begin{figure}
\plotfiddle{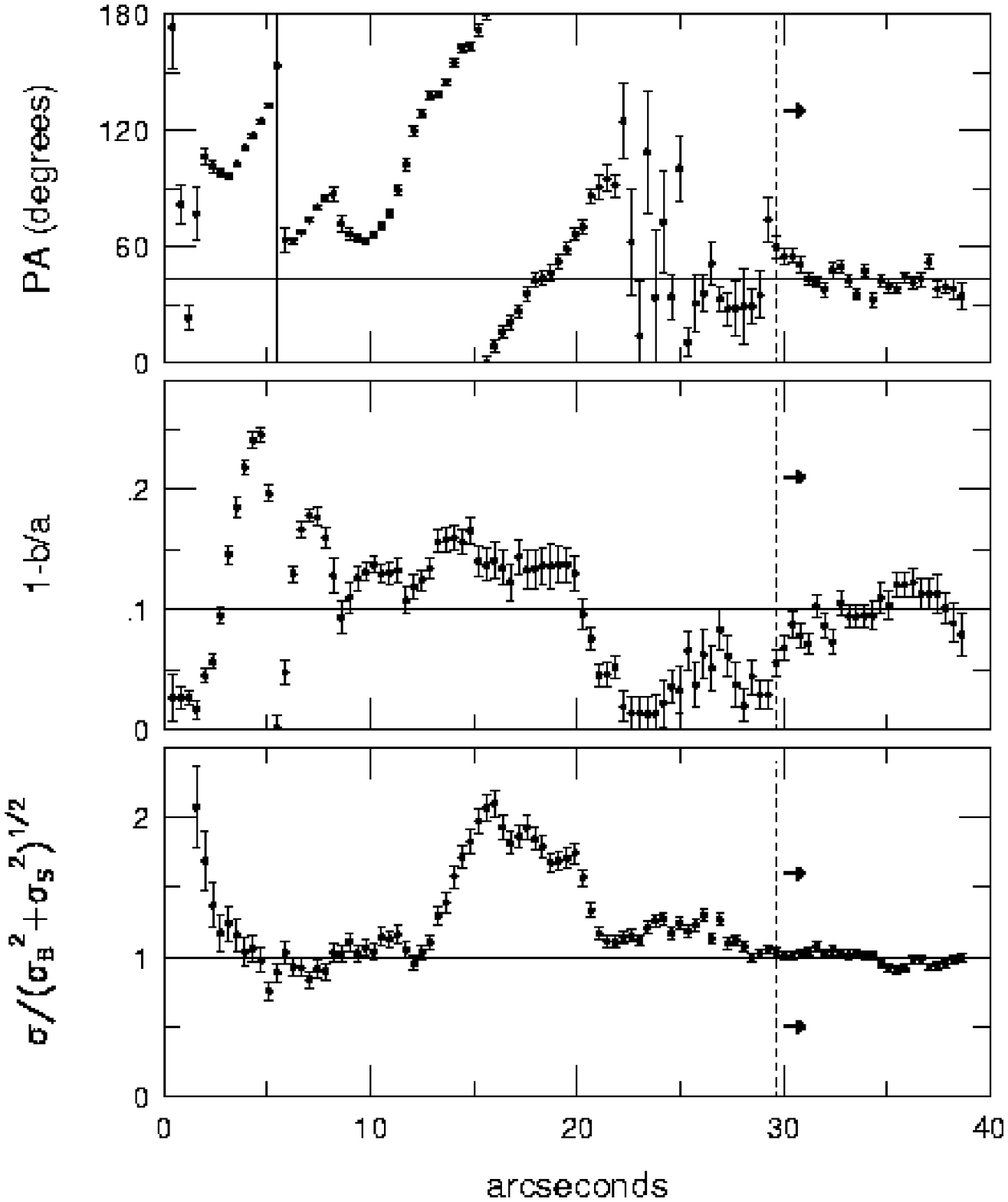}{5.65in}{0}{90}{90}{-300}{-155}
\end{figure}

\clearpage
 
\begin{figure}
\plotfiddle{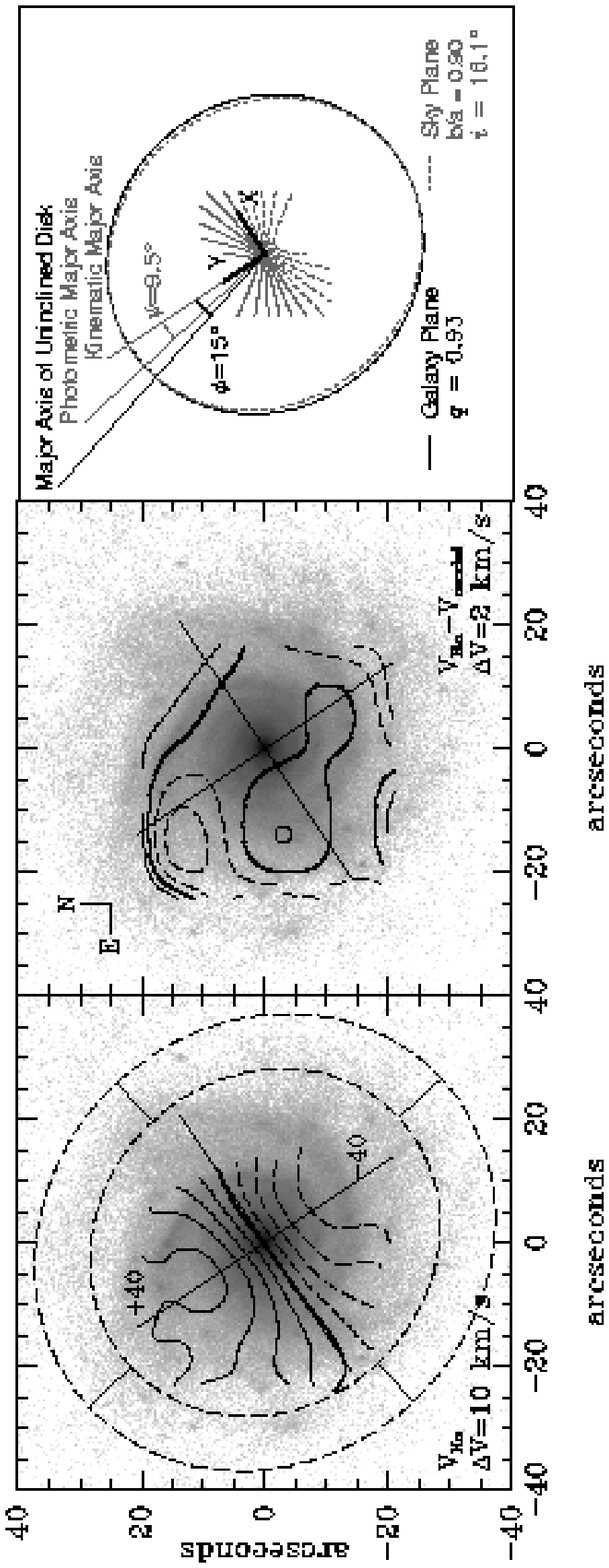}{5in}{0}{95}{95}{-285}{-185}
\end{figure}

\clearpage

\begin{figure}
\plotfiddle{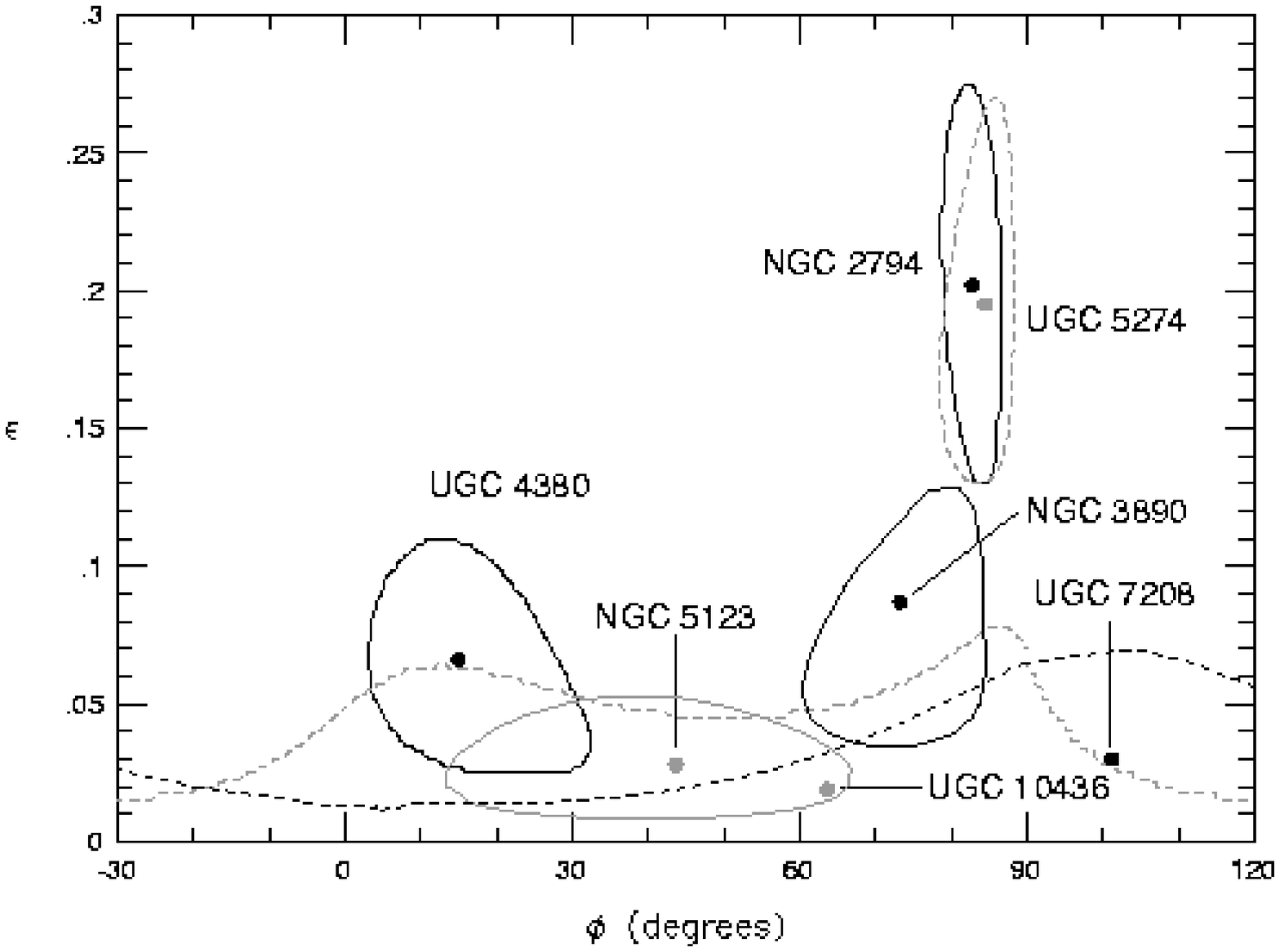}{5.65in}{0}{100}{100}{-310}{-175}
\end{figure}

\end{document}